\documentclass[floatfix,nofootinbib,twocolumn,showpacs,superscriptaddress, reprintnumbers,amssymb,letterpaper,amsmath,prl]{revtex4-2}

\usepackage{graphicx}
\usepackage{bm}
\usepackage[colorlinks=true,allcolors=blue]{hyperref}

\begin{document}

\newcommand{\mean}[1]{\mbox{$\langle{#1}\rangle$}}


\title{Observation of Skewed Electromagnetic Wakefields in an Asymmetric Structure \\
Driven by Flat Electron Bunches}

\author{W. Lynn}  
\email{wlynn@ucla.edu}
\affiliation{Department of Physics and Astronomy, UCLA, Los Angeles, California 90095, USA}
\thanks{These two authors contributed equally}

\author{T. Xu}
\email{xu@niu.edu}
\affiliation{Northern Illinois Center for Accelerator \& Detector Development and Department of Physics, Northern Illinois University, DeKalb, Illinois 60115, USA}
\thanks{These two authors contributed equally}
 
\author{G. Andonian}
\affiliation{Department of Physics and Astronomy, UCLA, Los Angeles, California 90095, USA}
\author{D. S. Doran}
\affiliation{Argonne National Laboratory, Argonne, Illinois 60439, USA}
\author{G. Ha}
\affiliation{Argonne National Laboratory, Argonne, Illinois 60439, USA}
\author{N. Majernik}
\affiliation{Department of Physics and Astronomy, UCLA, Los Angeles, California 90095, USA}
\author{P. Piot}
\affiliation{Northern Illinois Center for Accelerator \& Detector Development and Department of Physics, Northern Illinois University, DeKalb, Illinois 60115, USA}
\affiliation{Argonne National Laboratory, Argonne, Illinois 60439, USA}
\author{J. Power}
\affiliation{Argonne National Laboratory, Argonne, Illinois 60439, USA}
\author{J. B. Rosenzweig}
\affiliation{Department of Physics and Astronomy, UCLA, Los Angeles, California 90095, USA}
\author{C. Whiteford}
\affiliation{Argonne National Laboratory, Argonne, Illinois 60439, USA}
\author{E. Wisniewski}
\affiliation{Argonne National Laboratory, Argonne, Illinois 60439, USA}


\date{\today}

\begin{abstract}

Relativistic charged-particle beams which generate intense longitudinal fields in accelerating structures also inherently  couple to transverse modes.
The effects of this coupling may lead to beam break-up instability, and thus must be countered to preserve beam quality in applications such as linear colliders. 
Beams with highly asymmetric transverse sizes (flat-beams) have been shown to suppress the initial instability in slab-symmetric structures. However, as the coupling to transverse modes remains, this solution serves only to delay instability. In order to understand the hazards of transverse coupling in such a case, we describe here an experiment characterizing the transverse effects on a flat-beam, traversing near a planar dielectric lined structure.  The measurements reveal the emergence of a previously unobserved skew-quadrupole-like interaction when the beam is canted transversely, which is not present when the flat-beam travels parallel to the dielectric surface. 
We deploy a multipole field fitting algorithm to reconstruct the projected transverse wakefields from the data.
We generate the effective kick vector map using a simple two-particle theoretical model, with particle-in-cell simulations used to provide further insight for realistic particle distributions. 
\end{abstract}

\maketitle


%
%
%
%
%
%
%
%
%
%
%
%
%
%
%

Advanced acceleration concepts based on particle beam-driven wakefields are attractive candidates for next-generation accelerators, which aim to achieve high energy within a compact footprint~\cite{hidding-2019}.
A fundamental limitation of beam-driven wakefield acceleration in slow-wave structures and plasmas alike arises from the associated time-dependent transverse fields which induce beam breakup (BBU) instabilities  \cite{Li-2014}. 
Beam and structure configurations capable of circumventing such a limitation have been proposed to utilize symmetries other than axial in the structure and beam geometry. 
A simple implementation of such an approach consists of a slab-symmetric structure composed of two planar dielectrics surfaces, driven by a beam much wider than the gap between the dielectrics; this is referred to as a "flat" beam geometry ~\cite{tremaine-1997-a,Antipov:2012-apl, Andonian:2012}.
Theoretical~\cite{tremaine-1997-a, baturin-2018-a} and experimental~\cite{OShea:2020,Hoang:2018} results indicate that as the beam aspect ratio increases, the transverse field decreases at a stronger rate than the longitudinal accelerating field, thus permitting operation with lessened risk of BBU.
It has been previously known, on the other hand, that a normal quadrupole-like excitation exists in the flat beam driver scenario \cite{Li-2014}.
In this paper we examine this effect and also observe a newly uncovered phenomenon -- the excitation of a skew-quadrupolar wakefield which arises when the symmetry axes of the structure and beam are not parallel. 
To illuminate this scenario we present a simple model of the skew-quadrupole wakefield effect and report on experimental observations.  We then proceed to compare those results with the predictions from simulations and analytical models. 
We particularly concentrate on the experimental case of a single-sided structure, an unpaired slab where the skew-wakefield effects are enhanced, thus providing more clarity to our observations of the effects of the skew interaction.  To interpret the experimental results we develop a novel analysis algorithm to extract a transverse force field vector map from the projected beam distributions when the slab is present.
We also observe the presence of skew-wakefields in double-sided slab structures, similar to those used in wakefield accelerators or related terahertz radiation sources \cite{Andonian:2011,Antipov:2013}. 
Finally, we discuss the possible implications and applications of this emergent new class interaction. 

To gain some understanding of the skew-quadrupole wakefield excitation we develop an analytical model to provide an upper bound for the transverse wakefield forces associated with a single point-like charged particle propagating near a dielectric slab following the conformal-mapping approach discussed in Refs.~\cite{baturin-2016-a,baturin-2018-a}.
Specifically, Ref.~\cite{baturin-2016-a} shows that the resultant transverse force amplitude can be written as 
\begin{eqnarray} \label{eq:forceformula}
F_{\perp}(\omega,\omega_0)=F_0  \bigg[\frac{d^2 f(\omega,\omega_0)}{d\omega ^2}\bigg]^*\frac{df(\omega,\omega_0)}{{d\omega}},
\end{eqnarray}
 where $\omega\equiv x+iy$ describes the complex plane containing the structure, $\omega_0$ is the position of the source particle, and $F_0$ is a constant.
 The function $f(\omega, \omega_0)$ represents the conformal transformation that maps the geometry of the structure boundary onto a circle with the source located at its center (the symbol $^*$ indicates the complex conjugate). 
 Considering a single dielectric slab located at $y\le 0$ and assuming the upper plane $y>0$ to be vacuum, the M\"obius transformation $\omega\mapsto \chi(\omega)=\frac{\omega+i}{\omega-i}$ maps the upper plane to a disk with unit radius in the complex plane.
 Additionally, the transformation $g(\chi,\chi_0)=(\chi-\chi_0)/(1-\chi\chi_0^*)$, where $\chi_0\equiv \chi(\omega_0)$, ensures $\chi_0$ becomes the center of the disk. 
Thus, the net conformal map is given by the composition $f=(g\circ \chi )(\omega, \omega_0)$ and yields 
 \begin{eqnarray}\label{eq:confmap}
 f(\omega, \omega_0)= \frac{\left( \omega - \omega_{0}\right) \left({\omega_{0}^*} -  i\right)}{\left( {\omega_{0}}^* - \omega \right) \left(\omega_{0} +  i\right)}. 
 \end{eqnarray}
 Finally, the transverse force experienced by a test particle located at $(x,y)$ can be computed from Eq.~\ref{eq:forceformula} and \ref{eq:confmap} as   
 \begin{eqnarray}
F_{\perp} (x,y; x_0,y_0) =\frac{-2F_0}{[x_0-x+i(y+y_0)]^3}. 
\end{eqnarray}
The force components can be found from $F_{\perp}=F_x+iF_y$.  
\begin{figure}[hh!!!!!]
\includegraphics[width=1.0\columnwidth]{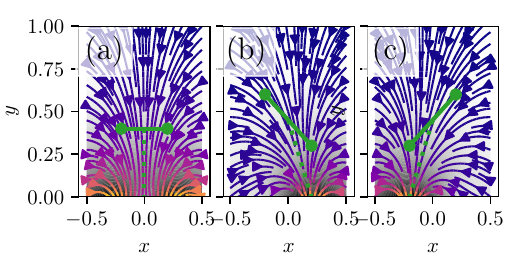}
\caption{\label{fig:theory_slabcases} Example of field lines associated with $F_{\perp}(x,y)$ for the case of two-particle beam (solid line with circles) injected parallel (a) or with negative (b) and positive (c) tilt with respect to the horizontal axis. The dashed lines indicate the directions of the force experienced by the center of the beam. }
\end{figure}
Figure~\ref{fig:theory_slabcases} illustrates the field lines associated with $F_{\perp}$ for the case of a flat beam, heuristically modelled as two horizontally-displaced particles. 
The force field lines in Fig.~\ref{fig:theory_slabcases}(a) show a pattern resembling the field lines of the normal component of a quadrupole magnet. 
When the two-particles are oriented with a tilt in the $x-y$ plane, the field map is altered dramatically, with a skewing of the field pattern towards the particle nearer the dielectric surface (Fig.~\ref{fig:theory_slabcases}(b,c)). 
The field pattern in this case is conceptually similar to the skew-component of a quadrupole magnet [$F_\perp\sim (x \hat{y}+ y \hat{x})$]. 
The flat beam scenario may be described through an extension of the heuristic two-particle model as it can be constructed out of a series of particles arrayed in linear fashion.

To extend the particle model, particle-in-cell simulations are employed to elucidate the effects of the transverse wakefields on realistic particle distributions.
The finite-difference time-domain programs {\sc warp}~\cite{Vay_2012} and {\sc CST particle Studio} ~\cite{cst} were used to study various configurations of planar-slab dielectric structures  and flat beam distributions such as those available at the Argonne Wakefield Accelerator (AWA)~\cite{xu-2021-a}. 
In these calculations, a  bi-Gaussian beam distribution with the experimentally-relevant parameters listed in Table~\ref{tab:slab-beam-param}  was propagated along a 15~cm dielectric structure, for both single-slab and double-slab configurations.

\begin{table}[hhhh!!]
\caption{Beam parameters measured at structure location.  \label{tab:slab-beam-param} }
\begin{center}
\begin{tabular}{l | c| c  c }\hline\hline
Parameter (Symbol) & \multicolumn{2}{c}{Value} & Unit \\
\hline
Charge ($Q$) & \multicolumn{2}{c}{$2.0\pm 0.3$} & nC  \\
Energy ($E$) & \multicolumn{2}{c}{$42 \pm 0.2$} & MeV \\
Horizontal emittance ($\varepsilon_{x}$) & \multicolumn{2}{c}{$196 \pm 19$} &\textmu{m} Rad\\ 
Vertical emittance ($\varepsilon_{y}$) & \multicolumn{2}{c}{$2.5 \pm 0.25$} & \textmu{m} Rad\\ 
rms bunch length ($\sigma_{z}$) & \multicolumn{2}{c}{$610\pm70$} & \textmu{m} \\
\hline
Tilt angle w.r.t. $x$-axis ($\theta$)  & 2.19 & 7.17& deg\\
rms horizontal size ($\sigma^*_{x}$) & $1.70\pm 0.01$ & $1.77\pm 0.01$  & mm  \\
rms vertical size ($\sigma^*_{y}$)   & $0.22\pm 0.02$ & $0.26\pm 0.01$  & mm  \\
\hline
\hline
\end{tabular}
\end{center}
\end{table}
%
%
The experiment to characterize the wakefield effects from a tilted flat-beam, as anticipated by the model and simulations introduced above, was performed at the AWA facility at Argonne National Laboratory~\cite{conde-2017-a}. 
The AWA beamline incorporates a radio-frequency photoinjector that produces a high-charge ($\sim 2$~nC) electron bunch, followed by a linear accelerator (linac) that boosts the beam energy to $\sim 42$~MeV. 
The photoinjector was operated in the ``blow-out" regime~\cite{musumeci-2008-a} by impinging a short (400~fs) ultraviolet laser pulse on the Cs$_2$Te photocathode.
In order to provide a flat-beam with high transverse aspect ratio, an axial magnetic field is applied at the cathode surface resulting in a beam with significant canonical angular momentum. 
As the beam exits the solenoid field, its canonical angular momentum components are fully transferred to mechanical angular momentum. 
The angular momentum is subsequently removed downstream of the linac by applying a torque using a set of three skew-quadrupole magnets. 
After this transformation, the transverse phase space is partitioned  such that the vertical phase-space area (quantified by its vertical normalized emittance $\varepsilon_y\equiv \frac{1}{mc}[\mean{y^2}\mean{p_y^2}-\mean{yp_y}^2]^{1/2}$ (where $\mean{...}$ indicates the averaging over the beam distribution) is much smaller than the horizontal emittance (\textit{i.e.} $\varepsilon_y \ll \varepsilon_x$)~\cite{piot-2006-a,xu-2021-a}.
This round-to-flat beam transformation procedure facilitates the creation of flat beams that achieve small vertical beam sizes at the waist while maintaining a long depth of focus. 

\begin{figure}
    \centering
    \includegraphics[width=\columnwidth]{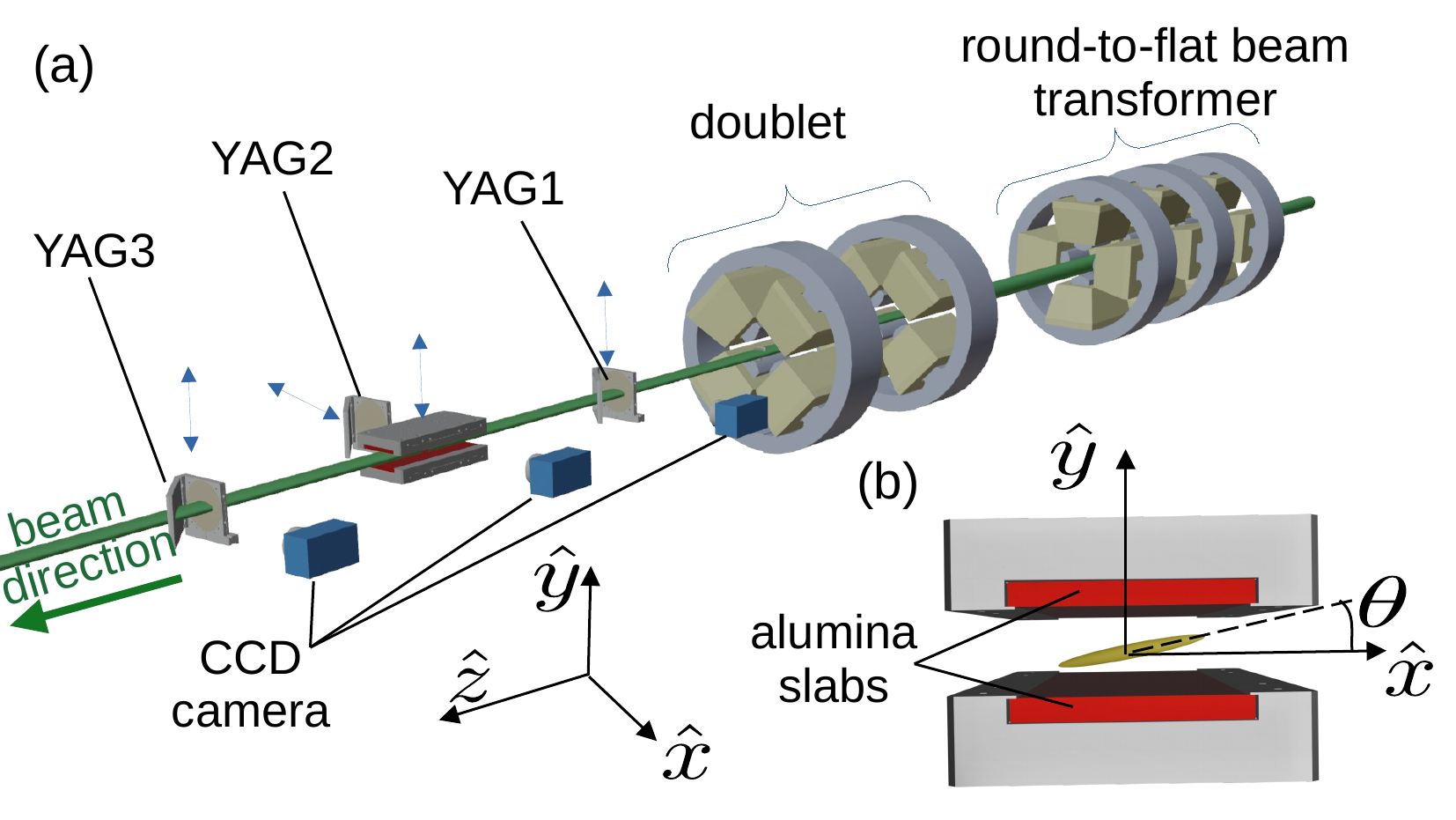}
    \caption{ (a) Layout of experiment in the vicinity of the structure, where the double-sided arrows indicate the direction of motions of remotely-controllable  devices, and (b) close-up view of the dielectric slab structure and flat-beam with incoming tilt angle, $\theta$. The two sides are independently controllable to vary the gap, and allow for single-slab studies. 
    \label{fig:layout}}
\end{figure}

The dielectric slab structure was installed downstream of a quadrupole magnet doublet which provides the final focusing  necessary to reduce the beam in size and transport it through the structure.
The dielectric structure, as described previously, is in a planar configuration consisting of two parallel slabs of alumina (Al$_2$O$_3$), with a relative permittivity of $\epsilon = 9.4$. 
The slabs are coated on one side with copper, forming an open waveguide with rectangular cross section as seen in the inset of Fig.~\ref{fig:layout}. This figure also depicts an overview of the key components of the AWA beamline used in the experiments.  
The main diagnostics in the experiment are Ce:YAG scintillating screens which are used in conjunction with CCD cameras to measure the transverse beam distributions, both before and after the structure.
The dielectric slabs have a transverse dimension of 4.9~cm, a thickness of 5~mm, and a length of 14.81~cm (along the beam  axis, or $\hat{z} $ direction). 
The slabs are mounted on a mobile support that allows for a variable vertical gap in the double-sided slab configuration, and full extraction of one slab for single-slab configuration. 
The measurements are parameterized by various beam injection angles and transverse offsets for both single- and double-sided configurations.
Alignment was performed with a helium-neon laser aligned collinearly with the beam centroid's trajectory and verified on the YAG1 and YAG3 screens (see Figure~\ref{fig:layout}). This permits coarse structure centering, with finer positioning achieved using a beam-based alignment method which entailed observing the vertical location that produced a minimal effect on the downstream beam distribution.

\begin{figure}[hh!!!!!]
\vspace{-5mm}
\begin{center}
    \includegraphics[width=1.0\columnwidth]{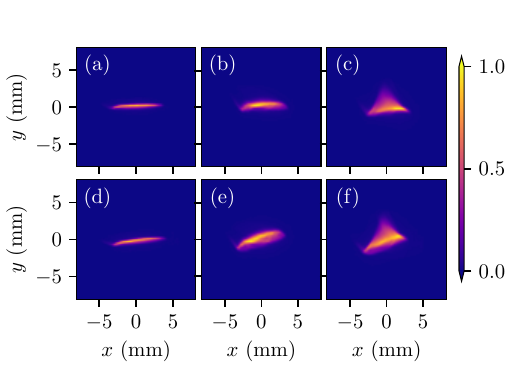}
\end{center}
\vspace{-5mm}
\caption{\label{fig:skew_wake_images} 
\textbf{Top Row:} Transverse beam images for injection angle of $\theta\approx~$2$^\circ$ measured at (a)  center of structure, YAG2, (b) downstream of structure without presence of slab at YAG3, and (c) downstream in presence of slab at YAG3.
\textbf{Bottom Row:} Transverse beam distributions for injection angle of $\theta\approx~$7$^\circ$ measured at (d) center of structure, YAG2, (e) downstream of structure without presence of slab at YAG3, and (f) downstream in presence of slab at YAG3.
All images use the same color map normalized to the peak value in image (a).  \label{fig:data-1slab}}
\end{figure}

Flat-beams were propagated through the structure offset from the dielectric surface by 0.91 mm to probe the dependence of the transverse field on the initial tilt angle, $\theta$, with respect to the $x-z$ plane. 
Image analysis using a machine vision technique based on the Hough transform \cite{illingworth1988survey, duda1972use}, is employed to compute $\theta$ from beam images recorded on the YAG screens. 
Measurements of the transverse distributions, for two different values of $\theta$, passing near a single-slab are depicted in Fig.~\ref{fig:data-1slab}. 
The flat-beam in the absence of the dielectric slab shows beam evolution consistent with free space propagation (Fig.~\ref{fig:data-1slab}(b)). 
In the case where the beam travels near the single-slab boundary, however, a clear transverse effect on the tail of the beam is apparent (Fig.~\ref{fig:data-1slab}(c)). 
As $\theta$ is increased, the transverse wakefields are altered, resulting in a skew in the field pattern and particle distribution at the tail of the beam (Fig.~\ref{fig:data-1slab}(f)).
The focusing effect is consistent with the analytical field pattern derived from the two-particle model depicted in Fig.~\ref{fig:theory_slabcases}.

\begin{figure}
    \centering
    \includegraphics[width=\columnwidth]{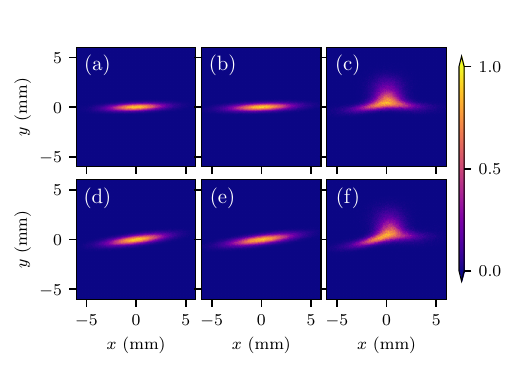}
    \caption{Simulations of transverse beam distribution at YAG2 (a,d) and YAG3 (b,c,e,f) with parameter correspondence to the data presented in Fig.~\ref{fig:data-1slab}.
    \label{fig:sim-1slab}}
\end{figure}

Particle-in-cell simulations for the measured beam parameters are performed to understand intra-bunch dynamics such as the kicks experienced by the beam tail particles. These simulations allow for a slice-by-slice analysis of the longitudinal dependence of the transverse kicks, information that complements the experimental observations of the time-integrated transverse beam profile. 
Figure~\ref{fig:sim-1slab} shows the transverse beam profiles for the same experimental parameters, integrated over all the slices of the beam from the {\sc warp} simulations.
The images show a pattern similar to that is Fig. ~\ref{fig:skew_wake_images} with quadrupole-like focusing effects observed on the tail of the beam when the bunch is injected parallel to the dielectric surface.
However, when the bunch is injected with an initial tilt of $\theta$=7$^\circ$, the wakefield is modified, and the tail particles experience kicks akin to those from a skew-quadrupole, in good agreement with the experimental data. 

\begin{figure}[hh!!!!!]
    \centering
\vspace{-2mm}
    \includegraphics[width=1.0\columnwidth]{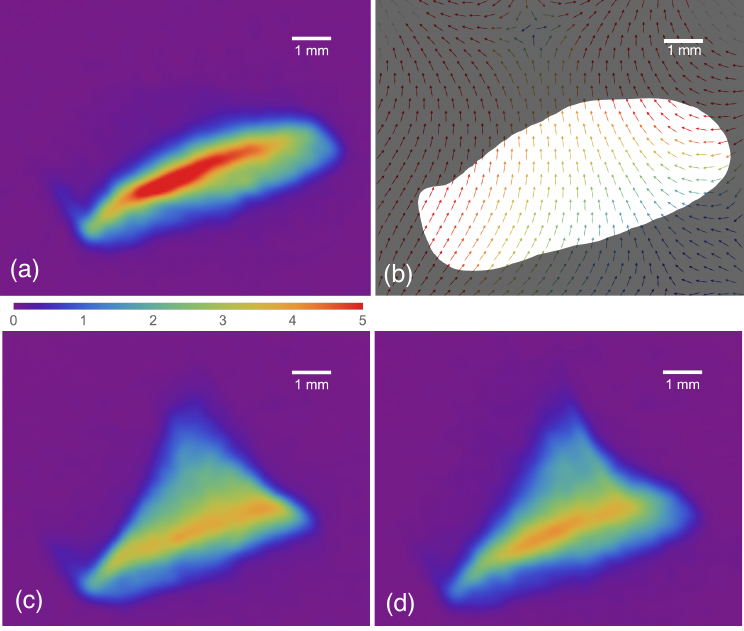}
\vspace{-5mm}
    \caption{\textbf{(a)} Beam transverse profile in absence of slab structure. (Charge density [arb. units] color-scale bar is constant between images). 
    \textbf{(b)} Reconstructed vector kick map. Unshaded region contains 90\% of beam charge. \textbf{(c)} Ground truth beam distribution after interaction with slab from experiment. \textbf{(d)} Implied final beam distribution after interaction from reconstructed wakefield kick map. 
    \label{fig:machinelearningfitting}}
\end{figure}

In order to gain further insight into the dynamics of the transverse beam evolution, a newly developed technique is used to reconstruct the wakefield forces from the available data ~\cite{majernik2023singleshot}. 
The algorithm accepts the projected images of the beam, with and without the dielectric boundary, as inputs. Then, using a recursive optimizer, the algorithm outputs the 2D spatial kick function as a vector field map.
An example of the process is shown in Fig.~\ref{fig:machinelearningfitting}. 
The initial beam distribution as measured on detector YAG2 (the incoming beam) is shown in Fig.~\ref{fig:machinelearningfitting}(a), along with the final beam distribution after interaction with the dielectric slab, as measured on detector YAG3  in Fig.~\ref{fig:machinelearningfitting}(c).
The integrated force (the kick, assuming no significant transverse motion in the dielectric region) map derived from the measured initial and final distributions is shown in Fig.~\ref{fig:machinelearningfitting}(b). 
The 2D kick map (Fig.~\ref{fig:machinelearningfitting}-b) generates the output distribution (Fig.~\ref{fig:machinelearningfitting}-(d)) which shows excellent agreement with the measured  final distribution.
Further examination of the kick map clearly shows the skew wakefield effect occurring in the beam region (unshaded area of Fig.~\ref{fig:machinelearningfitting}-(b)). 
The degree of skew in the wakefields is quantified by examining the skew and normal quadrupole coefficients of the kick map; for this case they are 0.31 and 0.24 respectively. 
The kick map generator is thus seen to be a powerful tool for revealing the projected transverse forces on the beam.

\begin{figure}[hh!!!!!]
\vspace{-5mm}
    \centering
    \includegraphics[width=\columnwidth]{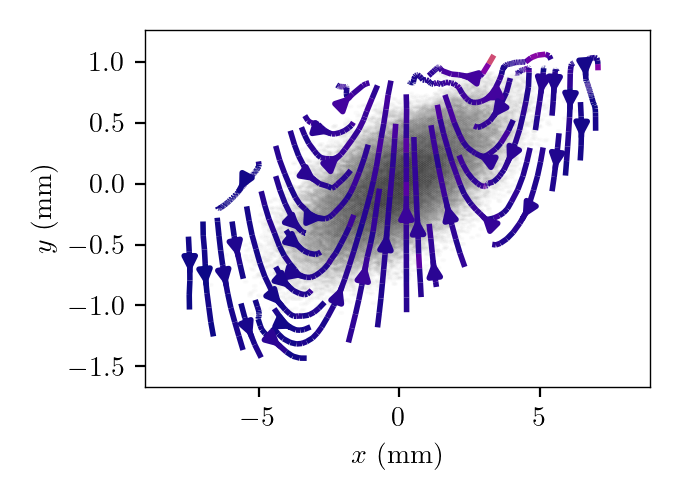}
\vspace{-5mm}
    \caption{Transverse field map fitted from particle momentum kicks in simulation for the same parameters as experiment.} 
    \label{fig:simulationfitting}
\end{figure}

The kick map reconstruction is compared to the simulation data of Fig.~\ref{fig:sim-1slab}.
The simulation kick map is generated by integrating the forces on the beam over the length of the structure, as shown in Fig \ref{fig:simulationfitting}. 
A comparison of the field map derived from simulations to that from the data shows agreement in the region of the beam.

\begin{figure}
\centering
\vspace{-5mm}
\includegraphics[width=\columnwidth]{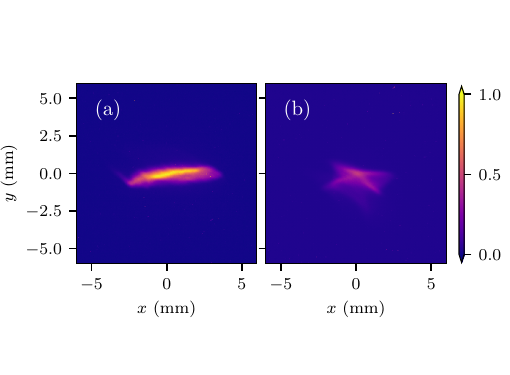}
\vspace{-10mm}
\caption{Experimental transverse distributions of the flat-beam with $\theta \approx 7^\circ$ (a)  without the structure present and  (b) after interaction through the two-sided structure. \label{fig:data-2slab}}
\end{figure}

The single slab measurements demonstrate the fundamental interactions with the skew wakefields for tilted beams. However, two-sided slab structures are also investigated due to relevance in applications such as advanced acceleration and phase-space manipulation.
As such, the experiment was repeated for the two-sided dielectric structure with a flat beam driver. 
In this scenario, the distance between the two parallel faces is 1.83 mm, and the flat beam was injected with the same $\theta$ as the single-slab case.
The main differences from the single-slab structure are the enhanced quadrupole and lessened dipole fields due to the presence of the second structure. 
An example of the double-slab data for the flat beam injected parallel to the two faces, and with $\theta \approx 7^\circ$ is shown in Fig.~\ref{fig:data-2slab}. 
The beam dynamics for this situation are intricate, as shown by the tail particles displaying a rotation not evident in the head of the beam. They require deconvolving of competing effects, which is subject of future study.




In the work presented here, flat-beams were used to interrogate details of the dielectric wakefield interaction. As flat beams strongly break rotational symmetry, when tilted they notably couple to skew transverse modes.  
The integrated wake-derived forces have been analyzed with a novel reconstruction tool that produces vector kick maps showing clear manifestation of a skew component of the transverse fields.
These transverse effects are important to characterize in applications, because if left uncontrolled, beam emittance dilution and instabilities (\textit{e.g.} short-range BBU) can arise.The newly examined considerations of skew interactions are important for dielectric and plasma wakefield accelerators, as well as for dielectric laser accelerators \cite{RevModPhys.86.1337}.
The single-slab studies are important for current applications in phase space manipulation such as dechirpers \cite{Bane:2016}, as well as in diagnostics exemplified by passive streakers \cite{Bettoni:2016}.

It should be noted that generally, wakefield focusing effects can also be employed to mitigate BBU, if employed in an alternating focusing gradient scenario. While in our study quadrupole-like fields are generated when finite-width flat beams are injected, use of symmetric (nearly round) beams in planar structures can purposefully excite the focusing fields which can be utilized to stabilize beam propagation over extended interaction lengths \cite{lynn2021strong}.

\begin{acknowledgments}
The authors acknowledge S. Baturin for conceptual development and insightful discussions.
This work is supported by the U.S. Department of Energy, Office of High Energy Physics, under Contracts DE-SC0017648 (UCLA), DE-SC0022010 (NIU), and DE-AC02-06CH11357 (ANL). 
\end{acknowledgments}

\bibliography{apssamp}

\end{document}